\documentstyle[prb,aps]{revtex}

\begin{document}

\twocolumn[\hsize\textwidth\columnwidth\hsize\csname
@twocolumnfalse\endcsname

\title{Spin Glass Behavior in $RuSr_{2}Gd_{1.5}Ce_{0.5}Cu_{2}O_{10-\delta }$}

\author{C.  A. Cardoso*, F. M. Araujo-Moreira}
\address{Grupo de Materiais e Dispositivos,  CMDMC, Departamento de F\'{i}sica, UFSCar,
Caixa Postal 676, 13565-905 , S\~{a}o Carlos-SP,
Brazil}
\author{V. P. S. Awana, E. Takayama-Muromachi}
\address{Superconducting Materials Center (Namiki Site), National Institute for
Materials Science,
1-1 Namiki, Tsukuba, Ibaraki 305-0044,
Japan}
\author{O. F. de Lima}
\address{Instituto de F\'{i}sica ''Gleb Wataghin'', UNICAMP, 13083-970, Campinas-SP,
Brazil}
\author{H. Yamauchi, M. Karppinen}
\address{Materials Science Laboratory, Tokyo Institute of Technology, Nagatsuta
226-8503, Yokohama,
Japan}

\date{Sept 24, 2002}
\maketitle

\begin{abstract}
The dynamics of the magnetic properties of polycrystalline $RuSr_{2}Gd_{1.5}Ce_{0.5}Cu_{2}O_{10-\delta }$ (Ru-1222) have been studied by ac
susceptibility and dc magnetization measurements, including relaxation and
ageing studies. Ru-1222 is a reported magneto-superconductor with Ru spins
magnetic ordering at temperatures near $100$ $K$ and superconductivity in
Cu-O$_{2}$ planes below $T_{c}\sim 40$ $K$. The exact nature of Ru spins
magnetic ordering is still debated and no conclusion has been reached yet.
In this work, a frequency-dependent cusp was observed in $\chi _{ac}$ vs. $T$
measurements, which is interpreted as a spin glass transition. The change in
the cusp position with frequency follows the Vogel-Fulcher law, which is
commonly acepted to describe a spin glass with magnetically interacting
clusters. Such interpretation is supported by thermoremanent magnetization
(TRM) measurements at T = 60 K. TRM relaxations are well described by a
stretched exponential relation, and present significant ageing effects.
\end{abstract}


\vskip1pc] \narrowtext

The coexistence of super\-con\-ductivity and magnetic order in ruthenium
copper oxides $Ru%
\mathop{\rm S}%
r_{2}(Gd,Sm,Eu)_{2}Cu_{2}O_{10-\delta }$ (Ru-1222) \cite
{Felner97,Felner98,Felner2002PRB65,Hirai02,Jardim02,Awana02} and $Ru%
\mathop{\rm S}%
r_{2}(Gd,Sm,Eu)Cu_{2}O_{10-\delta }$ (Ru-1212) has attracted a lot of
attention recently \cite
{Awana02b,Chen01,Bernhard99,Tallon99,Pingle99,Fainstein99,Lynn00,Williams,Tokunaga,Takagiwa,Chmaissem,Jorgensen,Butera}%
. But, besides this considerable interest, there are yet some unresolved
questions about the exact type of magnetic order in these compounds. The
difficulty about understanding the magnetic ordering in these systems is
that different techniques like muon spin rotation ($\mu $SR) \cite
{Bernhard99}, magnetic resonance (MR) \cite{Fainstein99}, neutron powder
diffraction (NPD) \cite{Lynn00,Takagiwa,Chmaissem,Jorgensen}, magnetization\ 
\cite{Williams,Butera} and nuclear magnetic resonance (NMR) \cite{Tokunaga},
though indicate towards canted antiferromagnetic ordering with a
ferromagnetic component, they do not agree completely with each other.
Although ferromagnetism and antiferromagnetism seems to be competing in
these compounds \cite{Lynn00,Takagiwa,Chmaissem,Jorgensen}, nobody
speculates about the possibility of this competition to cause a frustration
of the spin system, leading to a spin-glass scenario. The situation is
especially unclear for the Ru-1222 family. For Ru-1222, though NPD results
were reported recently \cite{Knee}, the magnetic structure has not been
unveiled. Although the magnetic behavior of Ru-1222 has been considered to
be analogous to the magnetic response for Ru-1212 samples, some recent
results point towards various differences \cite{Felner2002PRB65}.

In this work we explore the magnetic behavior of polycrystalline Ru-1222
samples by ac susceptibility $(\chi _{ac}=\chi ^{\prime }+i\chi ^{\prime
\prime })$, dc magnetization, and resistivity measurements. We observe a
significant dependence of $\chi _{ac}$ on the frequency of the excitation
field $h=h_{0}\sin (\omega t)$, which is characteristic of spin glass
systems. The temperature shift of the cusp in $\chi ^{\prime }$ follows the
Vogel-Fulcher law, which describes the freezing temperature $T_{f}$ of spin
glasses with magnetically interacting clusters. Such interpretation is
consistent with the observation of stretched exponential relaxations and the
occurrence of ageing effects at temperatures below the magnetic ordering
temperature.

\section{\bf Experimental details }

The $Ru%
\mathop{\rm S}%
r_{2}Gd_{1.5}Ce_{0.5}Cu_{2}O_{10-\delta }$ (Ru-1222) sample was synthesized
through a solid-state reaction route from $RuO_{2}$, $SrO_{2}$, $%
Gd_{2}O_{3}, $ $CeO_{2}$, and $CuO$. Calcinations were carried out on the
mixed powder at 1000, 1020, 1040, and 1060 $%
{{}^\circ}%
C$ each for 24 hours with intermediate grindings. The pressed bar-shaped
pellets were annealed in a flow of high-pressure oxygen (100 atm) at 420 $%
{{}^\circ}%
C$ for 100 hours and subsequently cooled slowly to room temperature \cite
{AwanaJLTP}. X-ray diffraction (XRD) patterns were obtained at room
temperature (MAC Science: MXP 18 VAHF$^{\text{22}}$; CuK$_{\alpha }$
radiation). Resistivity measurements were made in the temperature range of 5
to 300 $K$ using a four-point-probe technique. All ac susceptibility
measurements were performed in a commercial PPMS (Physical Properties
Measurement System), while for the dc measurements a SQUID magnetometer
MPMS-5 were employed, both equipments made by Quantum Design company.

\section{\bf Results and discussion }

Presently studied, Ru-1222 copper oxide sample crystallizes in a tetragonal
structure of space group $I4/mmm$ with $a=b=3.8327(7)$ \AA\ and $%
c=27.3926(8) $ \AA . The X-ray diffraction pattern, Fig. 1, shows a single
phase material, without any detectable impurity peak. The compound exhibited
superconductivity $(R\approx 0)$ below 40 $K$ in electrical transport
measurements \cite{AwanaJLTP}, as shown in the inset of Fig. 1. To recall
the characteristic magnetic behavior of $Ru-1222,$ Fig. 2 displays the
temperature dependence of both zero field cooled (ZFC) and field cooled (FC)
dc magnetization measured at $H=50$ $Oe$. The ZFC branch presents a
pronounced peak at $T_{p}=68$ $K$, just below the temperature where the ZFC
and FC curves separated. The freezing temperature $T_{f},$ extracted from ac
susceptibility measurements, is also indicated and will be discussed later.
At $T_{c}=45$ $K$ is observed a kink in both ZFC/FC curves as the $Ru-1222$
goes through its superconducting transition. The steady increase of the FC
branch at low temperatures is interpreted as being caused by the
paramagnetic response of the $Gd$ ions. It is important to notice that some
magnetic ordering starts to occur at a temperature $T^{\ast }=160$ $K$ much
higher than $T_{p}$. This can be observed in the inset of Fig. 2, which
shows an enlarged view of the magnetization curve at temperatures above $%
T_{p}$ revealing a small hysteresis at these temperatures. Interestingly,
both curves merge together again at a temperature around 80 K. The anomaly
observed at $T^{\ast }$ was previously reported as associated with an
antiferromagnetic transition \cite{Jardim02}, possibly in analogy to the
magnetic ordering occurring in $Ru-1212.$ In the same temperature region we
observed a small bump in $\rho \times T$ measurements (not shown). We
speculate if this anomaly could indicate the appearance of spin clusters
which at a lower temperature would have the magnetic moments frozen to
originate a spin-glass system. In Fig. 3 we present a magnetization curve
measured as a function of field, $M(H)$. The magnetization does not saturate
even at the highest field 90 kOe, as shown in the inset of Fig. 3, which is
consistent with what is expected for a spin-glass. The low field portion of
the virgin branch of the hysteresis loop at 60 K displays an S shape, with a
positive curvature at low fields, a typical characteristic of spin-glass
systems. It is important to notice that $T=60$ $K>>T_{c}=45$ $K$, thus this
positive curvature at low field is not due to a superconductor contribution
superimposed with a magnetic loop. Another striking characteristic of the
virgin branch is that it stays outside the hysteresis loop. This unusual
behavior was previously reported for cluster glasses with magnetic
interacting clusters \cite{Beck,Knitter,Zysler}. It seems to be related with
the more common displaced loop observed when the sample is field cooled \cite
{Beck,Knitter,Hurd,Ford}, though a more recent work consider it should be
due to a strong increase of the local surface anisotropy when the sample is
cooled below a certain characteristic temperature, for a system of nanosized
antiferromagnetic particles in an amorphous matrix \cite{Zysler}.

The ac susceptibility $(\chi _{ac})$ technique is a very powerful method to
provide evidence of a spin-glass behavior. In this case, both components $%
\chi ^{\prime }$ and $\chi ^{\prime \prime }$ of $\chi _{ac}$ present a
sharp, frequency dependent cusp. The position of the cusp in $\chi ^{\prime
} $ defines the freezing temperature $T_{f}$, which is coincident with the
temperature of the inflection point in $\chi ^{\prime \prime }$. It is also
well known that dc magnetic fields as low as a few hundreds of Oersted can
round this cusp up. In Fig. 4 we present the ac susceptibility for our
sample measured at $H_{dc}=50$ Oe. The main frame of Fig. 4 presents the
ZFC/FC temperature dependence of both $\chi ^{\prime }$ and $\chi ^{\prime
\prime }$ for the frequency $\nu =10000$ $Hz$. $\chi ^{\prime }$ presents a
sharp drop at the superconducting transition temperature $T_{c}$ and a
sharp, frequency dependent peak at $T_{f}\approx 72$ $K$. The peak shifts to
lower temperatures and its intensity increases as the frequency of the
excitation field is decreased (see upper inset, Fig. 4). For the $\chi
^{\prime \prime }$ peak we observe the shift to lower temperatures as well
as a decrease of its intensity with decreasing frequency (see lower inset of
Fig. 4). The frequency dependence of both components is a typical feature of
the dynamics of spin-glass systems. The coincidence of the temperature of
both, the peak in $\chi ^{\prime }$ and the inflection point in the $\chi
^{\prime \prime }$ curve, is also verified in our data. The $\chi ^{\prime }$
component present a double anomaly in the 110 - 170 $K$ range, but neither
frequency nor thermal-magnetic history dependences are observed. The
imaginary component does not present any significant feature in this
temperature range. On the other hand, for temperatures below 60 $K$ a clear
separation of the ZFC/FC curves is observed in both components, although it
is more prominent in $\chi ^{\prime \prime }$.

To further verify the existence of a spin-glass behavior, we have studied
the frequency dependence of $\chi _{ac}$ in more detail. A quantitative
measure of the frequency shift is obtained from $\Delta T_{f}/[T_{f}\log
(\omega )]$. This quantity varies in the range of 0.004 - 0.018 for
spin-glass systems \cite{Mydoshbook}, while for superparamagnets \cite
{Mydoshbook} it is of the order of 0.3. From a set of FC susceptibility
measurements at different frequencies, presented in the upper inset of Fig.
4, we could estimate $\Delta T_{f}/[T_{f}\log (\omega )]\approx 0.005$ for
our Ru-1222 sample. Therefore, our data are consistent with the spin-glass
hypothesis. There are basically two different possible interpretations of
the spin-glass freezing: the first one assumes the existence of a true
equilibrium phase transition at a finite temperature (canonical spin
glasses) \cite{Edwards}. The second interpretation assumes the existence of
clusters and, in this case, the freezing is a nonequilibrium phenomenon \cite
{Tholence}. For isolated clusters (superparamagnets), the frequency
dependence of their freezing temperature (in this context more correctly
referred as blocking temperature) has been predicted to follow an Arrhenius
law 
\begin{equation}
\omega =\omega _{0}\exp [-E_{a}/k_{B}T_{f}],  \label{Arrhenius}
\end{equation}
where $E_{a}$ is the potential barrier which separates two easy orientations
of the cluster and $\omega $ is the driving frequency of the $\chi _{ac}$
measurement. However, for magnetically interacting clusters, a Vogel-Fulcher
law has been proposed:

\begin{equation}
\omega =\omega _{0}\exp [-E_{a}/k_{B}(T_{f}-T_{0})],  \label{VF law}
\end{equation}
where $T_{0}$ can be viewed as a phenomenological parameter which describes
the intercluster interactions. Equation \ref{VF law} implies a linear
dependence of the freezing temperature with $1/\ln [(\omega \tau
_{0})^{-1}], $ $\tau _{0}=1/\nu _{0}=2\pi /\omega _{0}.$ In Fig. 5 we
present a Vogel-Fulcher plot, which shows that our data follows the expected
linear behavior. From the best linear fit we obtained $\nu _{0}\approx
1\times 10^{12}$ $Hz$, $T_{0}=66.92$ $K$ and $E_{a}=76.92$ $K$.

Also, the existence of the spin-glass behavior has been checked through the
time- dependent magnetic behavior of our sample. In this case,
thermoremanent magnetization (TRM) measurements were performed. Since the
behavior of a spin glass below $T_{f}$ is irreversible and complicated by
ageing processes, it is imperative to employ a well-defined H-T cycling
procedure to obtain meaningful data. The precise procedure adopted in this
work to measure the TRM relaxation was the following: the sample was field
cooled ($H=5000$ $Oe$) down from 200 K to 60 K; after temperature
stabilization we waited for a certain time $t_{w}.$ Thereafter the field was
reduced to zero and the magnetization was recorded as a function of the
elapsed time. The results for different values of $t_{w}$ $(100<t_{w}<1000$ $%
s)$ are presented in Fig. 6. Among the various functional forms that have
been proposed to describe the magnetic relaxation in spin glasses, one of
the most popular is the so-called {\it stretched exponential} 
\begin{equation}
M(t)=M_{0}-M_{r}\exp \left[ -\left( \frac{t}{t_{p}}\right) ^{1-n}\right] ,
\label{stretched exponential}
\end{equation}
where $M_{0}$ relates to an intrinsic ferromagnetic component and $M_{r}$ to
a glassy component mainly contributing to the relaxation effects observed.
Both $M_{r}$ and $t_{p}$ (the time constant) depend upon $T$ and $t_{w}$,
while $n$ is only a function of $T$. If $n=0$ one has the Debye, single
time-constant, exponential relaxation. On the other hand, for $n=1$, one
does not have any relaxation at all. The solid lines in Fig. 6 are the best
fits of equation \ref{stretched exponential} to our experimental data, with
parameters $4.38\times 10^{-3}<M_{0}<4.49\times 10^{-3}$ $emu$, $3.3\times
10^{-4}<M_{r}<3.7\times 10^{-4}$ $emu$, and $n=0.45$ (fixed for all fittings)%
$.$ The single parameter which presents a large variation with changes in
the wait time is the time constant $t_{p}$, which goes from $t_{p}=1749$ $s$
for $t_{w}=100$ $s$ to $t_{p}=5214$ $s$ for $t_{w}=1000$ $s$. The changes
observed in $M(t)$ measured for different values of $t_{w}$ demonstrate the
occurrence of ageing effects, what means that the physical system is in a
metastable state. In the inset in Fig. 7 this point is emphasized by showing
the relaxation rate $S(t)=\partial M/\partial \log (t)$. The shift in the
minimum position of $S(t),$ expected to occur for a spin-glass system, is
clearly observed.

\section{\bf Concluding remarks }

The frequency-dependent peak observed in the temperature dependence of the
ac susceptibility $\chi _{ac}$, combined with magnetic relaxation results,
provides strong evidence of the important role of magnetic frustration in
polycrystalline Ru-1222 to establish the existence of cluster glass
properties over a significant temperature range. This is to be contrasted
with the usual interpretation of the existence of long-range
antiferromagnetic order with spin canting for both $Ru-1222$ and $Ru-1212$
samples. The microscopic reason why $Ru-1212$ may present a long-range order
while $Ru-1222$ does not is not clear at this time. However, our results
come in line with the recent findings of \v{Z}ivkovi\'{c}{\it \ et al., }%
Ref. 3, who have pointed out significant differences in the magnetic
behavior of these two families of ruthenocuprates. Also, their results
indicate the existence of a metastable magnetic state below the magnetic
transition at $T_{f}$, which is in agreement with our interpretation of a
cluster glass freezing at $T_{f}$.

\section{\bf ACKNOWLEDGMENTS}

We thank L.M. Socolovsky for fruitful discussions. This work was supported
by Brazilian agencies Funda\c{c}\~{a}o de Amparo a Pesquisa do Estado de
S\~{a}o Paulo (FAPESP) through contracts \# 95/4721-4, 01/05349-4, and
Conselho Nacional de Pesquisas (CNPq) through contract \# 300465/88-2.

\label{aReferencias}

\vspace{0.5cm}

FIG. 1. Powder x-ray diffraction pattern for Ru - 1222 sample. Inset: $R(T)$
measurement at $H=0$ showing the superconducting transition.

\vspace{0.5cm}

FIG. 2. Field cooled and zero field cooled temperature dependence of
magnetization for $H=50$ $Oe$. Inset: amplification of the $M(T)$ curves
showing the small hysteresis at high temperatures.

\vspace{0.5cm}

FIG. 3. Low field portion of the $M(H)$ curve at $T=60$ $K$. Inset: entire $%
M(H)$ curve for fields up to 90 kOe.

\vspace{0.5cm}

FIG. 4. Complex susceptibility as a function of temperature for $\nu =10$ $%
kHz$ (main panel). Upper (lower) inset shows the frequency dependence of the
peak in the real (imaginary) component at the freezing temperature $T_{f}.$

\vspace{0.5cm}

FIG. 5. Variation of the freezing temperature $T_{f}$ with the frequency of
the ac field in a Vogel-Fulcher plot. The solid line is the best fit of Eq. 
\ref{VF law}.

\vspace{0.5cm}

FIG. 6. Thermoremanent magnetization relaxation for $T=60$ $K$ and $%
t_{w}=100 $ $s$ (squares)$,$ $500$ $s$ (circles), and $1000$ $s$
(triangles). The solid lines are the best fits of Eq. \ref{stretched
exponential}. Inset: relaxation rate $S(t)=\partial M/\partial \ln (t)$ for
the relaxations presented in the main panel.

\end{document}